# A Frequency Selective Surface based focal plane receiver for the OLIMPO balloon-borne telescope

Sumedh Mahashabde, Alexander Sobolev, Andreas Bengtsson, Daniel Andrén, Michael Tarasov, Maria Salatino, Paolo de Bernardis, Silvia Masi, and Leonid Kuzmin

*Abstract*—We describe here a focal plane array of Cold-Electron Bolometer (CEB) detectors integrated in a Frequency Selective Surface (FSS) for the 350 GHz detection band of the OLIMPO balloon-borne telescope. In our architecture, the two terminal CEB has been integrated in the periodic unit cell of the FSS structure and is impedance matched to the embedding impedance seen by it and provides a resonant interaction with the incident sub-mm radiation. The detector array has been designed to operate in background noise limited condition for incident powers of 20 pW to 80 pW, making it possible to use the same pixel in both photometric and spectrometric configurations. We present high frequency and dc simulations of our system, together with fabrication details. The frequency response of the FSS array, optical response measurements with hot/cold load in front of optical window and with variable temperature black body source inside cryostat are presented. A comparison of the optical response to the CEB model and estimations of Noise Equivalent power (NEP) is also presented.

## I. INTRODUCTION

Frequency selective surface is an artificial arrangement of simple metallic repeating patterns that provide a resonant interaction with incident electromagnetic wave. The interaction manifests as band-pass, band-stop, reflecting or absorbing behaviors [1]. A large number of simple repeating motifs have been used over the years in the form of crosses, squares, circles and hexagons, arranged either on planar substrate or as free standing metallic membranes that are used for quasioptical mesh filters [2]. The band-pass filter is usually a mesh design with narrow bandwidth and is a candidate for integrating bolometric detectors in. A number of attempts have been made to use FSS-based filters fabricated on membranes with a working principle that they couple to incident power in a given bandwidth heating the wideband sensitive membrane [3] [4] and the increase in the membrane temperature is detected by a Transition Edge Sensor. Even multi-color radiometers have been fabricated and tested based on this idea [4] [5].

The authors of this paper have been involved with the Cold Electron Bolometer Detector [6] which is a two terminal superconducting detector that uses two Superconductor-Insulator-Normal metal tunnel junctions to couple power to a normal metal nanoabsorber. The junctions are also used to measure the temperature change of the nanoabsorber. Most important feature of the CEB architecture is that the same two junctions cool of the electron system of the absorber using the bias current which provides negative electrothermal feedback. In this paper we describe detector architecture for the OLIMPO experiment [7]. In our planar pixel, we use the FSS approach for quasioptical coupling to incoming signal power from space and detection is provided using the CEB detector. The CEB is a two terminal detector and we've managed to arrange it into the repeating FSS motif e.g. annular shape, meander shape and match the impedance of the CEB to the embedding impedance seen by the CEB in such a way that RF currents flow through the CEB heating the electron system of the normal metal and creating a response signal. Such an integration of the planar imaging array with the FSS results in absorbing of the incoming radiation by the detector only in a narrow frequency band. The readout for this response uses the same FSS unit cells connected in series and parallel [6] [8] such that the Noise Equivalent Power (NEP) of the system together with the readout amplifiers is minimized.

The OLIMPO mission is a long duration balloon-borne experiment with a f/3.44 telescope and with detectors in four frequency bands cooled down to about 300mK. The goal of the experiment is photometric and spectrometric observations of the Sunyaev–Zel'dovich effect from galaxy clusters. The Band 3 of the telescope has 23 pixels in the focal plane. Commercial quasioptical mesh bandpass filters will be used to define the bandwidth from 330 GHz to 365 GHz. The detectors are expected to have a NEP to be lower than photon noise. The estimated background power loads are 38 pW and 66 pW in the photometric and the spectrometric configuration, respectively, which sets the background noise level. The

This work was supported by the Swedish National Space Board.

Sumedh Mahashabde, Andreas Bengtsson, Daniel Andrén Michael Tarasov and Leonid Kuzmin are with Chalmers University of Technology, Göteborg, Sweden

Alexander Sobolev and Michael Tarasov are with the Kotel'nikov Institute of Radio Engineering and Electronics, Moscow, Russia

Alexander Sobolev is also with Moscow Institute of Physics and Technology (State University), Moscow, Russia

Silvia Masi, Maria Salatino and Paolo de Bernardis are with University of Rome La Sapienza, 00185 Rome, Italy

readout for CEBs will be cold JFET amplifiers similar to [9] [10].

## II. FOCAL PLANE ARRAY OF DETECTORS

### A. FSS with integrated CEB detectors

We describe here the unit cell of the FSS structure the CEB detectors have been integrated in. It is an annular shape with extended lines on all 4 sides. One opposite pair of lines terminate at a short distance from the unit cell edge such that the other pair of lines can facilitate dc connections to neighboring cells in only one direction (X or Y). This pattern is designed to be fabricated on a Silicon substrate of calculated thickness backed with a backshort. The radiation is coupled from the Silicon side similar to traditional lens antennas. The FSS surface together with the substrate and backshort form a resonant interference filter that couples to submillimetre radiation, where the backshort and the substrate tune out the reactance of the FSS layer. This structure has been simulated in the commercial 3D electromagnetic simulation package HFSS using periodic boundary conditions and illuminated by a plane wave excitation using a Floquet port. The unit cell structure is shown in Fig. 1 with four integrated detectors in the arms of the annular element. The return loss ($S_{11}$) plot is shown in Fig 2. In our model the surface loss of the metal parts made of golden films was taken into account. Representing the four CEB-detectors each by a lumped port with the same impedance, we have found that the amount of power dissipated in the gold electrodes is negligible compared to the power absorbed in the detectors. So, the relation $\sum_{n=1}^{4} S_{1n}^2 + S_{11}^2 = 1$ is satisfied, where $S_{1n}$ is the power coupled to the n-th detector. Thus, $S_{11}$ can be used to describe the optical coupling to the detectors while simulating the complete pixel with hundreds of lumped elements.

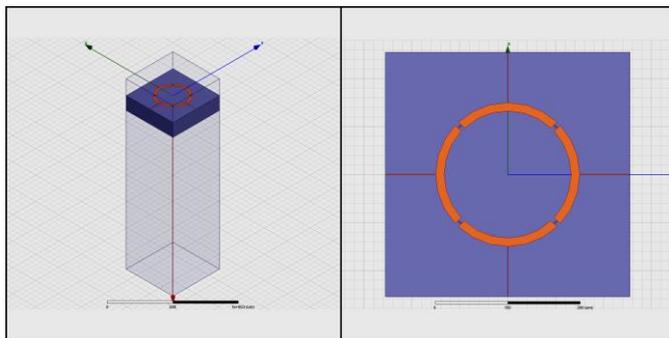

Fig 1. The unit cell of the FSS on the left and close up view of the annular mesh element on the right. The Floquet excitation is arranged at the bottom the unit cell. The grey solid is the silicon substrate with the annular shape visible on it. The top is capped by the backshort. Periodic boundaries are arranged on each parallel face to simulate an infinite array and the phase difference between the periods is used to calculate the angle dependent absorption of the FSS. The 4 bolometers (as four lumped elements) are arranged between the gaps of the annular element. The thin lines facilitate DC connection to neighboring pixels but are also an integral part of the FSS, being included in the RF simulations.

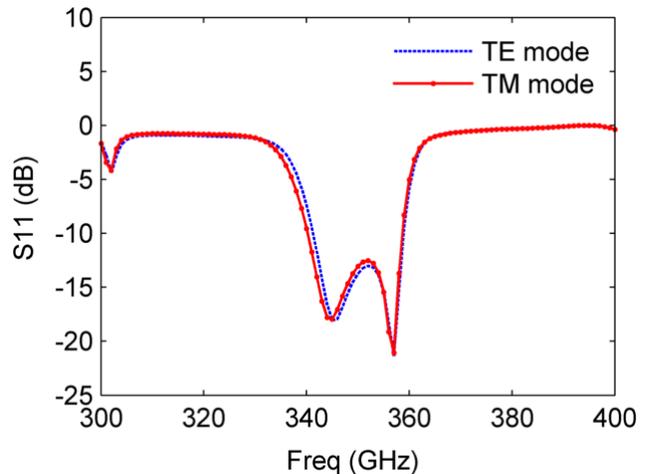

Fig 2. The $S_{11}$ plot of the FSS unit cell for two orthogonal modes of the Floquet port

The best case coupling has been calculated for a unit cell with a silicon substrate thickness of 141 μm, backshort distance of 160 μm and a unit cell pitch of 510 μm. The design has been optimized such that the best coupling is obtained for a real part of detector impedance of 50 Ω and an imaginary part of impedance corresponding to a capacitance of 50 fF. These are very common parameters in the CEB fabrication.

### B. Cold stop

Planar incoherent arrays of detectors can couple to incident radiation over a large solid angle and the advantages and disadvantages have been analysed in [11]. In the current design, we have made preliminary calculations of the absorption of an arbitrarily polarised wave incident on the unit cell at different angles. Once the return losses $S_{11}$(TE) and $S_{11}$(TM) for TE and TM modes are known from numerical simulations, we can estimate the angle dependent absorption of a finite pixel size by the formula P(TE,TM) = (1-$S_{11}^2$(TE,TM))cosθ. The factor cosθ accounts for the different pitch to the incoming radiation incident at angle θ relative to the focal plane normal vector. The result is shown in Fig.3. Here we neglected the edge effects and obtained the beam width 60° at the level – 3 dB. It is a narrower beam compared to that defined simply by cosθ function. The reason is that when the incident wave is not normal to the array plane (θ>0) the excitation signal has a phase delay across the unit cell that can be represented as an additional surface reactance. This reactance increases with θ and reduces the impedance match.

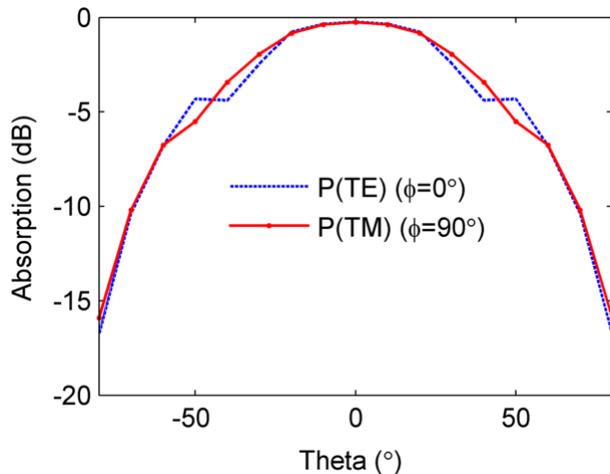

Fig 3. Angle dependent power absorption of the FSS for the two orthogonal (TE and TM) modes

For a pixel mounted in the cryostat of the OLIMPO telescope, the "sidelobes" should not illuminate any surface of the cryostat that could potentially radiate some power in the bandwidth of the filters. Due to thermal budget restrictions, a black absorbing surface nearest to the detectors could not be considered. Instead, a new design of an array of horn based cold apertures was developed to limit the angle of visibility of the pixel. These apertures would be mounted on the 300 mK stage of the cryostat atop the pixels. This aperture is shown in Fig 4, consists of two back to back horns connected together with a 2 mm long section of a circular waveguide of diameter 0.9 mm that is slightly overmoded (5 modes) at the frequencies of interest. The top part of the cold stop can be imagined as a traditional horn illuminating the telescope, and was simulated as such in the commercial package HFSS. The bottom part of the cold stop can be simulated by assuming a waveguide port at the circular waveguide end exciting all possible waveguide modes and propagating the power in the direction of the FSS pixel. The return loss ($S_{11}$) for each mode excited in the waveguide is a good indicator of the optical coupling of the pixel.

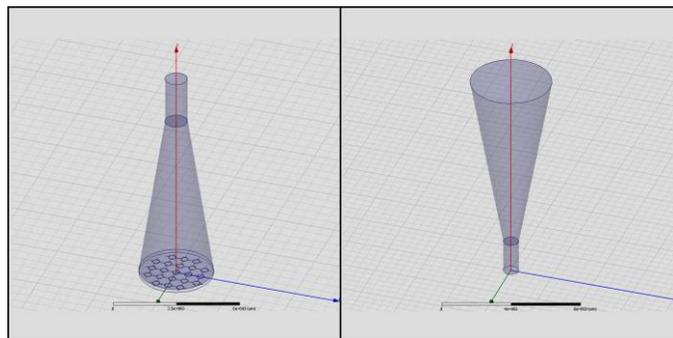

Fig 4. Model in HFSS of the lower part of the cold stop with CEB integrated FSS is shown on the left. Model of the upper part of the cold stop is shown on the right

We have simulated the cold stop + FSS in HFSS. The simulations of the top part of the cold stop indicate that it can illuminate the telescope optics reasonably well with an on-axis directivity of 23.4 dB and half power beam width of 12°. This directivity pattern is shown in Fig. 5 (top). The simulation of the bottom part of the cold stop with the FSS also shows promising results. For the full pixel of size ca. 3 mm x 3 mm, the optical coupling is better than 50% over the whole band. These results are shown in Fig. 5 (bottom).

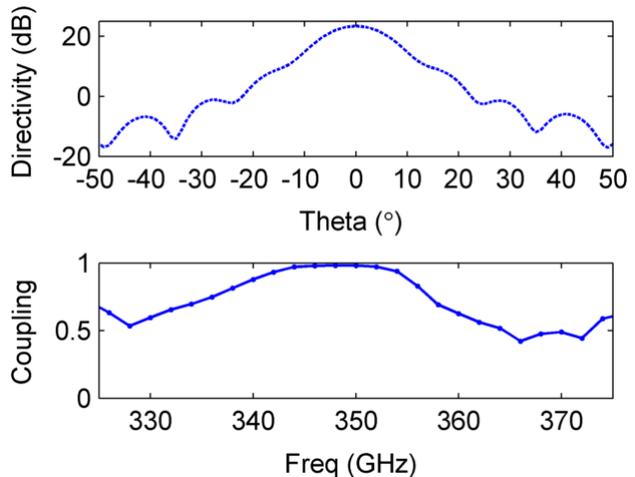

Fig 5. *(top)* Simulated directivity of the horn facing the telescope optics. *(bottom)* Simulation of the optical coupling of single pixel with cold stop

### C. DC design

The DC design for the integrated planar array follows from [12] [13]. Depending on the number of unit cells required to effectively cover the pixel area under the cold stop, the bolometers can be connected in series/parallel to match the noise impedance of the cold JFET amplifiers. In the current design, the cold stop limits the area of the pixel to a circle with a diameter of 3 mm where 21 unit cells of the annular shape can be fitted, each containing 4 bolometers, giving a total of 84 bolometers. The performance of the array can be simulated by modelling the performance of 1 CEB and scaling the responsivity to M array elements each experiencing M times less incident power. This analysis can be done using the heat balance equation [13] [14] [15] that equates power incoming into the normal metal absorber of the CEB and the power leaving the absorber due to bias current at quasi steady state.

$$2P_N + 2\beta P_S - \Sigma\Lambda(T_e^5 - T_{ph}^5) + I^2 R_a + \frac{P_0 + \delta P(t)}{M} = 0 \quad (1)$$

Here, $P_N$ is the power deposited in the normal metal due to bias current, $P_S$ is the power deposited in the superconducting electrode, $\beta$ is a parameter [16] that refers to the fraction of power deposited in the superconducting electrode that returns back to the normal absorber. $\Sigma\Lambda(T_e^5 - T_{ph}^5)$ is the heat flow from electron to phonon subsystems in the absorber, $\Sigma$ is the electron-phonon coupling constant which is material dependent, $\Lambda$ is the volume of the normal absorber, $T_e$ is the electron temperature of the absorber and $T_{ph}$ is the bath temperature. $I$ is the bias current, $R_a$ is the resistance of the absorber. $P_0$ is the background power on the pixel in steady state while $\delta P(t)$ is the incident signal power. Assuming that the signal power is a very tiny variation on top of the background power (this is especially true for balloon borne experiments with very large atmospheric power loads), the pixel dc parameters can be estimated using the steady state part of the heat balance equation.

The current through a SIN junction is dependent on the number of occupied and unoccupied states in each of the metals and can be expressed as [14]

$$I = \frac{1}{eR_N} \int_{-\infty}^{\infty} \nu(E)\big(f_N(E-eV) - f_S(E)\big)dE \quad (2)$$

where $e$ is the electron charge, $R_N$ is the normal resistance of the junction and $V$ is the voltage across the same junction. $\nu(E)$ is the Dynes density of states [17] for the superconductor and $f_N(E)$ is the Fermi-Dirac distribution for the normal metal which depends not only on the energy but also on the temperature of the electron subsystem in the normal metal. This dependence of temperature is what we correlate to absorbed power in the normal metal. $f_S(E)$ is the Fermi-Dirac distribution for the superconductor and in general it will be different from $f_N(E)$ since they can have different electron temperatures.

The tunneling electrons in the SIN junction will carry with them some power; this power $P_N$ can be calculated by removing one $e$ in (2) and replace it with $eV - E$ which is the energy deposited in the normal metal from one tunneling event. Doing this we obtain [14]

$$P_N = \frac{1}{e^2 R_N} \int_{-\infty}^{\infty} (eV-E)\nu(E)\big(f_N(E-eV) - f_S(E)\big)dE \quad (3)$$

It is possible for $P_N$ to be negative and this is the effect we use to cool the normal metal, giving the detector its name.

From energy conservation we conclude that the power dissipated in the superconductor is

$$P_S = IV - P_N \quad (4)$$

The responsivity of the CEB in current-bias mode is given by

$$S_v = -2 \frac{\frac{\partial I}{\partial T}/\frac{\partial I}{\partial V}}{G_{e-ph} + 2G_{SIN}} \quad (5)$$

Here $G_{e-ph}$ is the heat conductance due to electron-phonon coupling and $G_{SIN}$ is the heat conductance of one tunnel junction [15]. $\frac{\partial I}{\partial T}$ and $\frac{\partial I}{\partial V}$ are the partial derivatives of the current with respect to temperature and voltage.

The total NEP of the system can be divided into three parts, the SIN-junction itself, the electron-phonon interaction and also the amplifier noise. The squares of the three parts can be calculated as

$$NEP_{SIN}^2 = \delta P^2 - 2 \frac{\delta P \delta I}{\frac{\partial I}{\partial V}S_v} + \frac{\delta I^2}{\left(\frac{\partial I}{\partial V}S_v\right)^2} \quad (6)$$

$$NEP_{e-ph}^2 = 10 k_B \Sigma \Lambda (T_e^6 + T_{ph}^6) \quad (7)$$

$$NEP_{amp}^2 = \frac{V_n^2 + \frac{N}{W}\left(I_n^2\left(\frac{\partial V}{\partial I} + R_a\right)\right)^2}{\left(\frac{S_v}{W}\right)^2} \quad (8)$$

In the SIN junction there exist three different sources of noise; $\delta P^2$ is the heat flow noise, $\delta I^2$ is the fluctuations in the current or shot noise and since the current and the heat flow are correlated there also exists a correlation term $\delta P \delta I$ [15]. The amplifier contributes two noise sources, $V_n$ the voltage noise and $I_n$ the current noise. Assuming $N$ is the number of bolometers in series and $W$ is the number of bolometers in parallel, giving a total of $N*W$ bolometers, the total NEP is then given by

$$NEP_{total}^2 = N*W*NEP_{SIN}^2 + N*W*NEP_{e-ph}^2 + NEP_{amp}^2 \quad (9)$$

The most optimized combination of these bolometers taking into account the geometry and lowest possible NEP over the full range of incident powers (20 pW to 80 pW, with a margin on both sides of OLIMPO estimations) is 42 bolometers in series and 2 in parallel. This corresponds to 21 unit cells of the FSS.

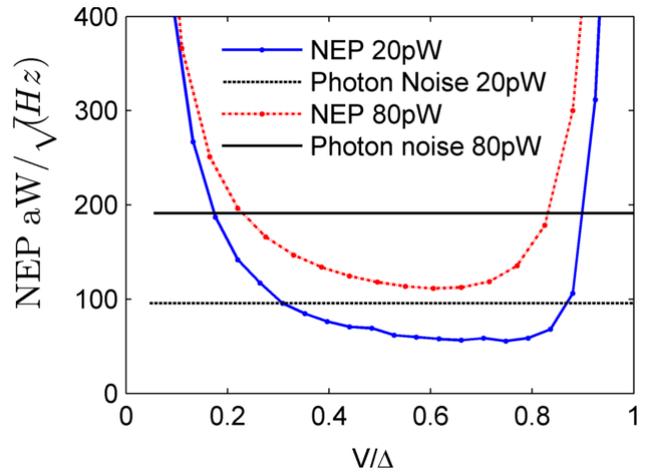

Fig 6. Contributions of noise to the pixel design with the voltage scale normalized to one junction in the array for power load of 20 pW and 80 pW. This plot has been calculated for $R_a = 150\ \Omega$, Rn = 1 kΩ, Volume of absorber $\Lambda = 0.023\ \mu m^3$, $\beta = 0.1$, $\Sigma = 1.5\ nW/(K^5 \mu m^3)$. Amplifier Voltage noise $V_n = 3nV/\sqrt{Hz}$ and current noise $I_n = 3fA/\sqrt{Hz}$

The NEP plots for two different power loads calculated from equations 1-6 are shown in Fig. 6. We notice from the plots that the pixel with 42x2 bolometers performs well with NEP of the device less than the photon noise over a range of powers from 20 pW to 80 pW. Thus, the same pixel can be used for both the photometric and spectrometric configurations, whence, in the spectrometer configuration the background power is a lot larger owing to emissions of uncooled spectrometer mirrors.

### D. Fabrication

The fabrication process is similar to the one described in [18]. The pixel is fabricated using multiple lithography layers. The initial layer consists of alignment marks and defines the pixel size. The next layer is used to expose and evaporate contact pads and the FSS structure. Silicon machining using deep Silicon etch process constitutes the third layer. The Silicon substrate is thinned down to the required thickness using the Bosch process. The Bosch process used for deep silicon etching uses the STS® ICP plasma etcher with $SF_6$ gas in the etch step and $C_4F_8$ gas in the passivation step providing an anisotropic etch of the silicon. The process is stopped at the required depth with an etch error typically less than 2 µm. The final layer is the exposure of the CEB devices in an electron beam lithography machine. The size of each of the two bolometer SIN tunnel junctions is 2 µm x 0.4 µm connected with the nanoabsorber being 2 µm long and 100 nm wide and 14 nm thick. After exposure and development of the e-beam resist, the bolometer layers are evaporated. The normal metal absorber is a layer of aluminum whose superconductivity is suppressed by a thin layer of ferromagnetic impurities (Fe). The tunnel barrier of the SIN tunnel junction is formed by oxidation of the normal metal at 10mbar oxygen pressure for 5 minutes. The superconducting electrodes are then evaporated at two angles (+45° and −45°) to a thickness of 70 nm each. Finally, all unwanted metal and resist is lifted off using warm acetone and a gentle ultrasonic agitation.

*E. Prototype testing*

The first prototype pixel was fabricated and characterized at low temperatures. The test consisted of measuring voltage response of the device in front of open cryostat windows with room temperature and liquid nitrogen loading. The incoming radiation was filtered with quasioptical bandpass filters along with Fluorogold to decrease infrared power. The optical response to 300 K/77 K loading is shown in Fig. 7. The IV curves of the device are shown on the left axis; the curve with lower linearity corresponds to 300 K radiation. The optical response is plotted on the right axis as the difference in voltage of the two IV curves. This test highlights the optical sensitivity of the device. The power incident on the device with the optical load at 300 K is more than 550 pW and at 77 K is more than 150 pW. Assuming the optical coupling from Fig. 5, the device is operating beyond its designed power load and is overheated. This causes the responsivity to be far lower than at design power loads; here estimated about $4.5 * 10^6 \frac{V}{W}$.

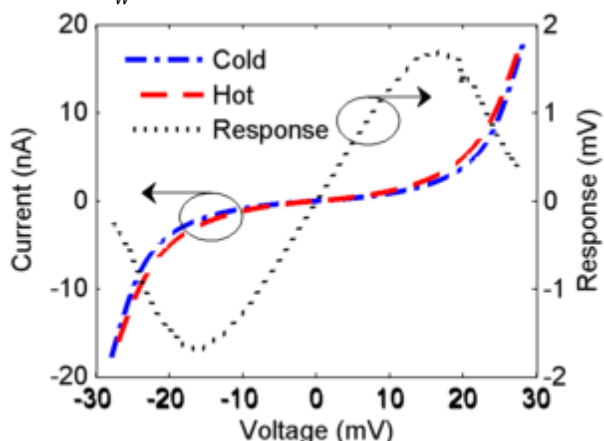

Fig. 7. Optical response with 300K and 77K loading

For testing the device at lower power loads, a test setup with a cold black body source shown in Fig. 8 was used. The prototype was connected to the 300 mK stage of a cryostat. The sample holder had blackened inner surface to reduce spurious reflections and lowpass and bandpass filters to define the correct optical band. The RF source was a cold black body with a heater and thermometer connected to the 2.7 K stage of the cryostat using a dielectric anchor to prevent heat leaks to the cryostat shields. The incident power on the pixel could be varied by changing the black body temperature between 2.7 K and 6 K.

The incident power could be estimated using Planck's law. Current-Voltage curves and Voltage-Power curves of the CEB detectors were recorded by changing the black body temperature.

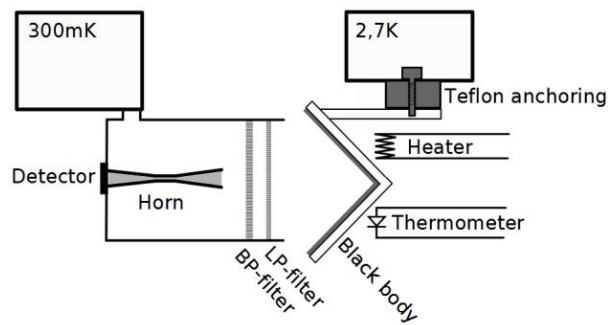

Fig. 8. The prototype measurement setup

Measurement of the response of the pixel is shown in Fig. 9. The estimated responsivity is also shown in the figure; this is calculated using estimated incident power on the device. The responsivity decreases with increasing incident power which is due to increasing electron temperature of the absorber. In this calculation, the black body emissivity was assumed to be 0.9 and the blackened inner surface of the sample holder is assumed to attenuate any stray radiation falling on it. The fluorogold cover to the sample holder was assumed to have a transmission of unity in the bandwidth of interest.

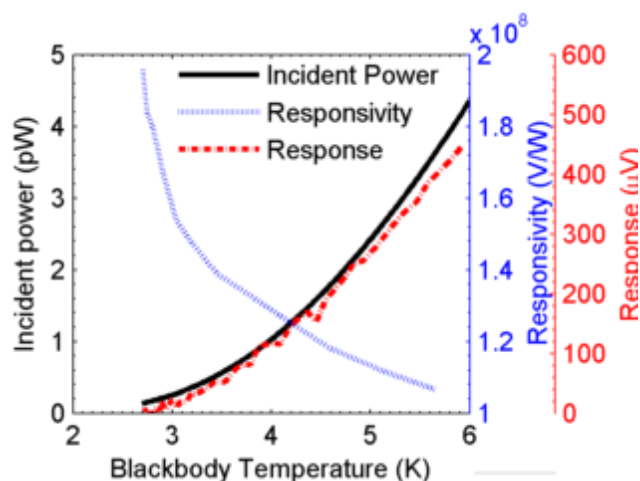

Fig. 9. Incident power, optical response and estimated responsivity of the pixel

Another way to estimate the responsivity of the device is by fitting the optical response data to the bolometer model presented in equations 1-5. An IV curve of the device is taken with the black body kept at a known temperature. This data is then fitted to the model described earlier and responsivity is calculated. For small changes in radiated power from the black body, this responsivity curve can very reasonably describe the change in absorbed power in the bolometer array. Thus the value of absorbed power can be estimated. Fig. 10 shows data fits to optical response taken at black body temperatures of 2.7 K, 3.6 K and 4.5 K. The two response curves are generated by subtracting individual IV curves at one black body temperature from the lowest one. The data fits indicate that about 0.2 pW of power was absorbed at 3.6 K and about 1 pW power was absorbed by the array at 4.5 K black body temperature. Compared with Fig 9, the estimated absorbed power is lower by about a factor of 2, and the estimated responsivity correspondingly higher. The difference can be attributed to over estimation of emissivity of the black body, under estimation of attenuation of one of the elements in the optical path, difference in the frequency response of the array with the cold stop compared to simulations, or a combination of these instances.

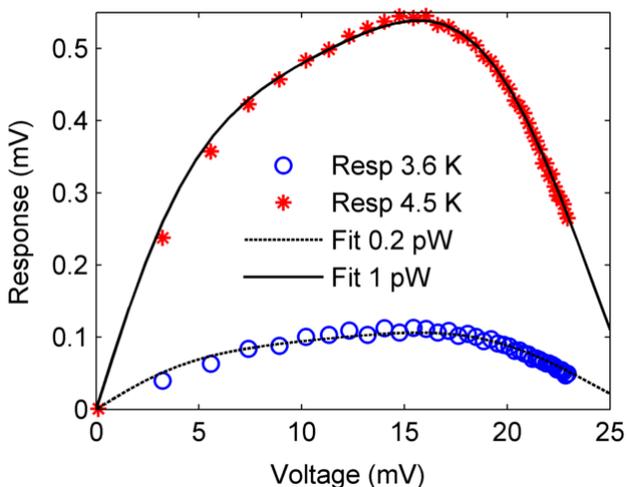

Fig. 10. Optical response of the array at two black body temperatures and data fits for absorbed power from the CEB model. This fit has been calculated for $R_a = 150\ \Omega$, Rn = 2 k$\Omega$, Volume of absorber $\Lambda = 0.023\ \mu m^3, \beta = 0.03, \Sigma = 1.1\ nW/(K^5\mu m^3)$.

Using the above estimated responsivity, one can also estimate the electrical NEP by dividing the measured value of noise (at 115 Hz) by the responsivity. This is presented in Fig 11. The background power is estimated to be 6 pW from the data fit and the corresponding photon noise level is indicated. Using the equations 6-9, the calculated NEP for the actual parameters of the fabricated sample is also indicated. The current and voltage noise values are referred to the commercial JFET amplifier AD743 beyond its 1/f corner. This amplifier was used for noise measurement in the actual experiment.

The estimated NEP is about $1.5 * 10^{-16}\ \frac{W}{\sqrt{Hz}}$ at optimum bias point and is about 4 times higher than the modelled one. At higher absorbed power, the dynamic resistance of the array will be lower and the impact of the current noise of the JFET amplifier will be reduced. The measurement showed low frequency interferences that were the likely source of the increased noise and of the 6 pW background power indicated earlier. One source of the low frequency interference is the vibrations of the pulse tube cooler of the cryostat used in the experiment. It was not possible to switch it off to reduce its effects since it introduced temperature gradients during the course of the measurement.

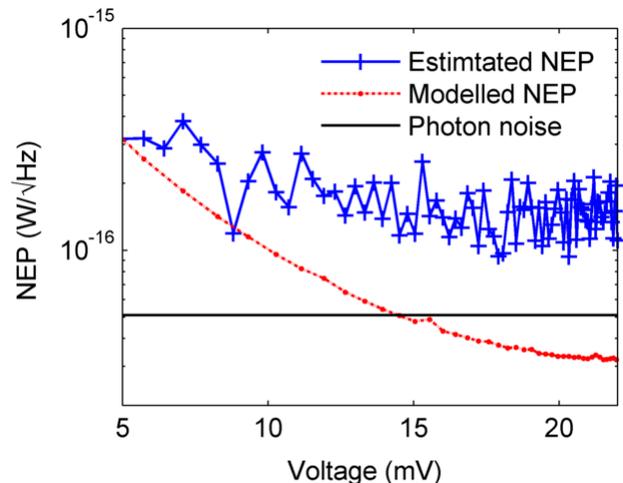

Fig. 11. Comparison of estimated NEP, modelled NEP and photon noise. Amplifier voltage noise $V_n = 4nV/\sqrt{Hz}$ and current noise $I_n = 5fA/\sqrt{Hz}$

To understand the frequency response of the FSS, an experiment was made where the FSS was illuminated with radiation from a Backward Wave Oscillator (BWO) through a bandpass filter chain. Initially the filter chain was characterized separately and then used in the measurements of the FSS. Given the variation of BWO emitted power versus the operating frequency we normalized the measured voltage with the corresponding power reporting the normalized efficiency in the radiation detection. The results are shown in Fig. 12. The filter frequency response has a large enough bandwidth to cover the FSS band. The FSS itself has a frequency response comparable to the simulations shown in Fig. 2.

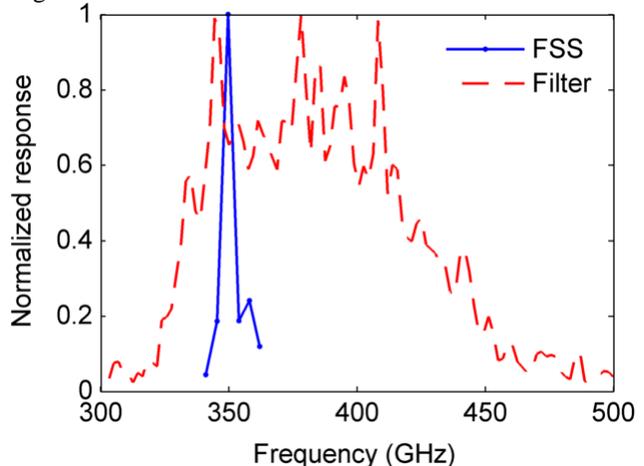

Fig. 12. BWO measurement of the frequency response of the FSS and the bandpass filter characteristics.

III. DISCUSSION

Compared to the use of traditional coupling approaches like spider-web bolometer or the slot antennae, the distributed FSS array is conducive to the use of two terminal CEB devices. In the case of the spider-web or the slot antennae, a single (or few) detector absorbs power that is coupled to the spider-web or through the immersion dielectric lens that usually accompanies the slot antenna. In the case of the FSS array, incident power is distributed over a number of bolometers such that the array of CEBs acts as a distributed absorber along with a distributed coupling element (the FSS geometry). This distribution of power causes each bolometer to be "cooler" which improves the responsivity. In the case of OLIMPO, it was not possible to design a single CEB element that could handle the specified power, so the FSS concept was developed.

The CEB detectors could be used in large-scale arrays, which are needed for almost all future instruments. The device fabrication is not an issue since facilities and competence exists to fabricate arrays over the size of up to 6-inch wafers (> hundred pixels). The most important consideration is the readout system. The multiplexing readout technology for medium/high-impedance sensors is relatively undeveloped. Analog multiplexers based on Si-Ge ASICs could solve the problem [19], by means of low-noise MOSFET switches operating at low temperature with low power dissipation, but this has not been tested with the arrays. Another solution is low-ohmic arrays matched to Superconducting Quantum Interference Device (SQUID) readout. For this goal, well developed SQUID readout systems with multiplexing could be used [20]. For matching of relatively high-Ohmic CEB with low-ohmic SQUID a parallel array of CEBs was proposed for realization of background-limited operation [8]. If impedance matching of the CEB array and SQUID is not enough an additional superconducting transformer could be effectively used for matching.

IV. CONCLUSION

We have designed and fabricated a single pixel of a submillimetre focal plane receiver using Frequency Selective Surface based arrays integrated with the Cold Electron Bolometer detector. Optical response was measured with the cold stop designed for the OLIMPO project and responsivity of $2*10^8 \frac{V}{W}$ was estimated at low power loads and $4.5*10^6 \frac{V}{W}$ at high power loads. Electrical NEP for this array has been estimated to be $1.5*10^{-16} \frac{W}{\sqrt{Hz}}$ at low background power. These results of the prototype testing indicate very promising characteristics of the pixel. We hope to use this approach for future submillimetre focal plane based pixel arrays.

ACKNOWLEDGEMENTS

The authors acknowledge the help of Lars Jönsson in the fabrication of cold stop and the sample holder and of Alessandro Schillaci and Giuseppe D'Alessandro in the FTS measurements of the filters chain transmission.

REFERENCES

[1] B. A. Munk, *Frequency selective surfaces: theory and design*. John Wiley & Sons, 2005.

[2] P. A. R. Ade, G. Pisano, C. Tucker, and S. Weaver, "A review of metal mesh filters," in *Astronomical Telescopes and Instrumentation*, 2006, p. 62750U–62750U.

[3] M. Kowitt, D. Fixsen, A. Goldin, and S. Meyer, "Frequency selective bolometers," *Applied optics*, vol. 35, no. 28, pp. 5630–5635, 1996.

[4] T. Perera, T. Downes, S. Meyer, T. Crawford, E. Cheng, T. Chen, D. Cottingham, E. Sharp, R. Silverberg, F. Finkbeiner, and others, "Optical performance of frequency-selective bolometers," *Applied optics*, vol. 45, no. 29, pp. 7643–7651, 2006.

[5] A. Datesman, J. Pearson, G. Wang, V. Yefremenko, R. Divan, T. Downes, C. Chang, J. McMahon, S. Meyer, J. Carlstrom, and others, "Frequency selective bolometer development at Argonne National Laboratory," in *Astronomical Telescopes and Instrumentation: Synergies Between Ground and Space*, 2008, pp. 702029–702029.

[6] L. Kuzmin, "Optimization of the Hot-Electron Bolometer and A Cascade Quasiparticle Amplifier for Space Astronomy," in *International Workshop on Superconducting Nano-Electronics Devices*, 2002, pp. 145–154.

[7] F. Nati, P. Ade, A. Boscaleri, D. Brienza, M. Calvo, S. Colafrancesco, L. Conversi, P. de Bernardis, M. De Petris, A. Delbart, and others, "The OLIMPO experiment," *New Astronomy Reviews*, vol. 51, no. 3, pp. 385–389, 2007.

[8] L. Kuzmin, "Distributed Antenna-Coupled Cold-Electron Bolometers for Focal Plane Antenna," *Proc. ISSTT*, pp. 154–158, 2008.

[9] B. Crill, P. A. Ade, D. Artusa, R. Bhatia, J. Bock, A. Boscaleri, P. Cardoni, S. Church, K. Coble, P. de Bernardis, and others, "Boomerang: A balloon-borne millimeter-wave telescope and total power receiver for mapping anisotropy in the cosmic microwave background," *The Astrophysical Journal Supplement Series*, vol. 148, no. 2, p. 527, 2003.

[10] T. E. Montroy, "Measuring CMB polarization with BOOMERANG," New Astronomy Reviews, 47, Issue 11-12, 1057-1065, 2003.

[11] M. J. Griffin, J. J. Bock, and W. K. Gear, "Relative performance of filled and feedhorn-coupled focal-plane architectures," *Applied Optics*, vol. 41, no. 31, pp. 6543–6554, 2002.

[12] L. Kuzmin, "Two-dimensional array of cold-electron bolometers for high-sensitivity polarization measurements," *Radiophysics and Quantum

*Electronics*, vol. 54, no. 8–9, pp. 548–556, 2012.

[13] L. Kuzmin, "An array of cold-electron bolometers with SIN tunnel junctions and JFET readout for cosmology instruments," in *Journal of Physics: Conference Series*, 2008, vol. 97, no. 1, p. 012310.

[14] G. C. O'Neil, "Improving NIS Tunnel Junction Refrigerators| Modeling, Materials, and Traps," 2011.

[15] D. Golubev and L. Kuzmin, "Nonequilibrium theory of a hot-electron bolometer with normal metal-insulator-superconductor tunnel junction," *Journal of Applied Physics*, vol. 89, no. 11, pp. 6464–6472, 2001.

[16] P. Fisher, J. Ullom, and M. Nahum, "High-power on-chip microrefrigerator based on a normal-metal/insulator/superconductor tunnel junction," *Applied physics letters*, vol. 74, no. 18, pp. 2705–2707, 1999.

[17] R. Dynes, V. Narayanamurti, and J. P. Garno, "Direct measurement of quasiparticle-lifetime broadening in a strong-coupled superconductor," *Physical Review Letters*, vol. 41, no. 21, pp. 1509–1512, 1978.

[18] M. Tarasov, L. S. Kuzmin, V. S. Edelman, N. Kaurova, M. Y. Fominskii, and A. Ermakov, "Optical response of a cold-electron bolometer array," *Jetp Letters*, vol. 92, no. 6, pp. 416–420, 2010.

[19] F. Voisin, E. Bréelle, M. Piat, D. Prêle, G. Klisnick, G. Sou, and M. Redon, "Very low noise multiplexing with SQUIDs and SiGe heterojunction bipolar transistors for readout of large superconducting bolometer arrays," *Journal of Low Temperature Physics*, vol. 151, no. 3–4, pp. 1028–1033, 2008.

[20] P. Ade, R. Aikin, M. Amiri, D. Barkats, S. Benton, C. Bischoff, J. Bock, J. Brevik, I. Buder, E. Bullock, and others, "BICEP2 II: Experiment and Three-Year Data Set," *arXiv preprint arXiv:1403.4302*, 2014.

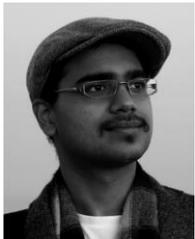
**Sumedh Mahashabde** was born in India in 1985. He received his B.E degree in Instrumentation Engineering from Mumbai University, India, in 2007 and M.Sc. in Microtechnology from Chalmers University of Technology, Gothenburg, Sweden in 2010. He is currently pursuing his Ph.D. in Physics from Chalmers University, Sweden. His research interests include terahertz bolometers, superconducting tunnel junctions and Graphene.

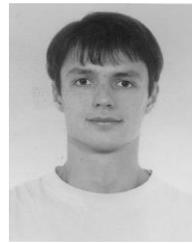
**Alexander Sobolev** obtained M.Sc. and Ph.D. degrees from Moscow Institute of Physics and Technology (State University) in 2003 and 2006 respectively. His master and Ph.D. research was related to the development of the fully superconducting integrated receiver for the instrument TELIS. During this work he made a number of research visits to the Technical University of Denmark (DTU) and Netherlands Institute for Spece Research (SRON). At present, Alexander Sobolev is a senior research fellow at Kotel'nikov Institute of Radio Engineering and Electronics (Russian Academy of Sciences). His current research interests include long Josephson junctions, submm detectors and heterodyne receivers, integrated superconducting circuits, submm antennas and focal plane detectors array.

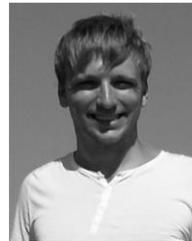
**Andreas Bengtsson** was born in Falkenberg, Sweden, in 1991. He received the B.Sc. degree in engineering physics in 2013 from Chalmers University of Technology, Sweden where he is currently studying towards a M.Sc. degree in nanotechnology. His research interests include terahertz bolometers, superconducting tunnel junctions, transmon qubits and superconducting microwave resonators.

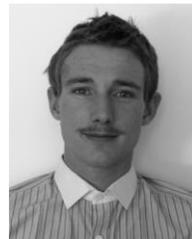
**Daniel Andrén** received the B.Sc. degree in Engineering Physics from Chalmers University of Technology, Sweden in 2013. During the subsequent part of the year he worked in the Bolometer Group in the Quantum Device Physics laboratory with Cold Electron Bolometers for the OLIMPO project. Daniel will embark on his M.S. Degree in Nanotechnology in 2014.

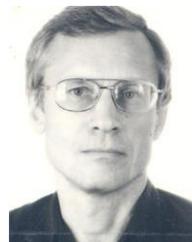
**Mikhail A. Tarasov** received the master degree from M. Lomonosov Moscow State University in 1977, the Ph.D. degree in 1983, and the Doctor of Sciences (Habilitation) degree in 1997 from V. Kotel'nikov Institute of Radio Engineering and Electronics of Russian Academy of Sciences. Since 1977 he has been with V.Kotel'nikov Institute, where he is currently a Principal Investigator. For the last 20 years, he has been spending up to six months every year with Chalmers University of Technology, Gothenburg, Sweden, collaborating in development of SIS mixers, Josephson detectors, SQUID amplifiers, and cold electron bolometers. The list of his publications includes over 200 papers in scientific journals and proceedings of international conferences. His research interests are mainly in superconducting electronics, low temperature physics, microwave spectroscopy, and noise in superconducting devices.

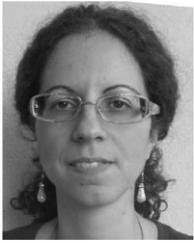

**Maria Salatino** was born in Rome, Italy, in 1983. She took both her M.Sc. cum laude in Astronomy and Astrophysics in 2007 and Ph.D. in Astronomy in 2011 from University of Rome La Sapienza. Her research interests include cold electron bolometers, cryogenic polarization modulators and systematic effects in far-infrared polarimeters. She is involved in projects of the Italian (LSPE), the French (PILOT) and the European (ESA ITT) space agencies. She is a referee for the Journal of Low Temperature Physics.

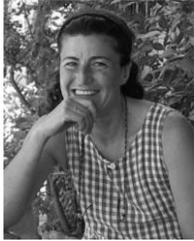

**Silvia Masi** received her Ph.D. in Physics in 1984. She is now associate professor at the University of Rome La Sapienza. She works developing advanced instruments to study the Cosmic Microwave Background. She has been in charge of the cryogenic system for several CMB instruments, is a member of the Planck collaboration, and is the PI of the OLIMPO balloon-borne telescope to study the SZ effect

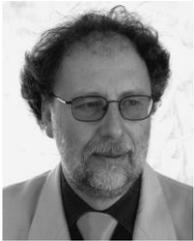

**Paolo de Bernardis** received his Ph.D. in Physics in 1984. He is now professor of astrophysics at the University of Rome La Sapienza. His research focuses on the experimental study the Cosmic Microwave Background using bolometric detectors. He has coordinated several balloon-borne experiments, has been co-PI of the BOOMERanG experiment, and is a member of the Planck collaboration.

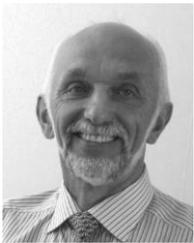

**Leonid S. Kuzmin** was born in Moscow, Russia, in 1946. He received the Ph.D. degree in physics (Degree of Candidate of Science in Physics and Mathematics) from Moscow State University, Moscow, in 1977, with the thesis topic "Nondegenerate single-frequency parametric amplification using Josephson junctions with self-pumping" defended in 1997 and "Correlated Tunnelling of Electrons and Cooper Pairs in Ultrasmall Tunnel Junctions."
In 2000, he was a Docent with Chalmers/Göteborg University. Since 2009, he has been a Professor with Chalmers University of Technology, Gothenburg, Sweden. He is the Chairman of 13 international workshops and three inter-national schools for young scientists with a general title "From Andreev Reflection to the Earliest Universe." He has authored more than 200 publications (including 114 in referred journals; citation index: 1305, h-index: 19, average citations per item: 11.45)

.